\documentclass[preprint,3p]{elsarticle}
\usepackage{graphicx}
\usepackage{graphics}
\usepackage{enumitem}
\usepackage{amssymb}
\usepackage{color}

\journal{Applied Energy}

\begin{document}

\begin{frontmatter}

\title{An Electromagnet-Based Magnetically-Activated Thermal Switch Without Moving Parts}

\author{C. Rodrigues\textsuperscript{1}}
\author{M. M. Dias\textsuperscript{1}}
\author{L. Martins\textsuperscript{1}}
\author{J. P. Ara\'ujo\textsuperscript{1}}
\author{J. C.R.E Oliveira\textsuperscript{2}}
\author{A. M. Pereira\textsuperscript{1}}
\author{J. Ventura\textsuperscript{1}}
\address{\textsuperscript{1}IFIMUP and IN-Institute of Nanoscience and Nanotechnology, Departamento de F\'isica e Astronomia da Faculdade de Ci\^encias da Universidade do Porto, Rua do Campo Alegre 687, 4169-007 Porto, Portugal}
\address{\textsuperscript{2}CFP, Department of Engineering Physics, FEUP, R. Dr. Roberto Frias s/n, 4200-465 Porto, Portugal}

\begin{abstract}

With the ever increasing power dissipation in electrical devices, new thermal management solutions are in high demand to maintain an optimal operating temperature and efficient performance. 
In particular, recently developed magnetically-activated thermal switches (MATSs) provide an alternative to existing devices, using the magnetic and thermal properties of superparamagnetic nanofluids to dissipate heat in a controlled manner.
However, the presence of moving parts is a major drawback in those systems that must still be addressed.
Herein, we present a compact and automatized MATS composed by an encapsulated superparamagnetic nanofluid and an electromagnet allowing to activate the MATS without any moving part.
We investigate the effect of different temperature gradients ($40$, $26$ and $10$ $^{\circ}$C) and powers applied to the coil (6.5, 15, 25 and 39 W) on the performance of this novel MATS. 
The results show that the highest ($44.4\%$) and fastest ($0.6$ $^{\circ}$C/s) temperature variation occur for the highest studied temperature gradient. 
On the other hand, with increasing power, there is also an increase in the efficiency of the heat exchange process between the two surfaces.
These results remove one of the main barriers preventing the actual application of magnetic thermal switches and opens new venues for the design of efficient thermal management devices.

\end{abstract}


\end{frontmatter}

\section{Introduction}

The scavening for new and improved ways of controlling a system's temperature and managing heat transfer between objects is an important topic, with the thermal management market growing $7.6\%$ annually \cite{market2014}. With the development of novel microelectromechanical systems (MEMS) arrives the need for thermal management solutions capable of adapting to different operational modes and changes in the environment's temperature. One of the solutions for this issue lies on the use of a thermal switch (TS) \cite{Puga2017,Gou2014,Cottrill2015,Wehmeyer2017}. A TS allows controlling the transfer of heat between two surfaces by establishing and breaking a thermal connection between them. This ON/OFF capability allows the thermal switch to adapt and perform differently depending on the desired operating mode and design.

Typically, TSs can be divided into three main categories: active solid-state, passive solid-state and fluidic thermal switches \cite{Kitanovski2014}. Active thermal switches need to be externally activated by applying a voltage \cite{Marland2004,Jahromi2014}, magnetic \cite{Peshkov1965,Bartlett2010,Chen1997,Samantaray2015,Sato1993,Jeong2012a} or electric fields \cite{Yeom2008}, among other \cite{Zhu2014}. This generates or breaks contact between the heat source and the cold surface. This is also their main disadvantage, as the thermal contact resistance between surfaces can lead to significant heat losses.

Passive solid-state thermal switches, or thermal diodes, function with materials that transfer heat asymmetrically, meaning that, for the same temperature, the heat flow for one direction is larger than for the opposite direction. This eliminates the need for an external activation mechanism and can be achieved by pairing two materials with different thermal conductivities, one that varies with temperature and the second nearly constant for the same temperature range. These pairs can be made of polymers \cite{Pallecchi2015}, metals \cite{Starr1936,Geng2011} or other materials \cite{Chang2006}.

However, the interface between two solids offers limitations to thermal transport usually caused by surface roughness that reduces the actual contact area \cite{Cha2009}. Although interface engineering, such as the use of carbons nanotubes \cite{Xu2006}, can mitigate the issue, the main problem still remains.
Fluids remove this problem altogether: by adapting to the contact surface, the interface resistance is decreased.
Thus, fluidic thermal switches have been widely reported in the literature \cite{Cho2007a,DiPirro2014,Kimball2012,Jeong2013,Badar2016}.
Liquid metal coolants have been used taken advantage of the electrowetting effect \cite{Ma2007}. 
Nanofluids \cite{Murshed2016}, made of high thermal conductivity nanoparticles suspended in a fluid, display enhanced thermal properties and are more effective in heat removal. 
In particular, magnetic nanofluids, besides offering enhanced thermal conductivities, can be externally redirected to specific areas on demand by just applying small magnetic fields \cite{Puga2017,Badar2016}.
Thus, they can be moved into localized hot spots where they can absorb heat that is then released at a far away location after the field is removed.
However, up to now, magnetic nanofluids have been primarily used as a secondary transport mechanism rather than a primary mean of heat transportation \cite{Leland1991,DOSHISHA;STELLAGREENCORP2012}.

In a previous work \cite{Puga2017}, the authors reported a novel magnetically-actuated thermal switch (MATS) relying on the mechanical motion of permanent magnets to move a magnetic nanofluid between two surfaces at different temperatures. 
Such apparatus allowed the realization of heat transference between hot and cold sides with a remote operation (Fig. \ref{TS2}).
However, and although with a reliable and efficient performance, the necessity for moving parts is a majority disadvantage for the majority of actual applications, preventing its use in a broad range of systems.
Taking this into consideration, we idealized and developed a new design in which all mobile parts are removed and that relies instead on the magnetization/demagnetization processes of a soft magnetic material.
With the reported results, we provide the groundwork for future applications of the MATS.
This device can also provide a future alternative to the more commonly used phase change materials \cite{Wang20151,Mortazavi2017,Wu2017,Wang2017} in thermal management applications.

\section{Experimental Details}

\begin{figure}[!t]
    \centering
    \includegraphics[width=3.5in]{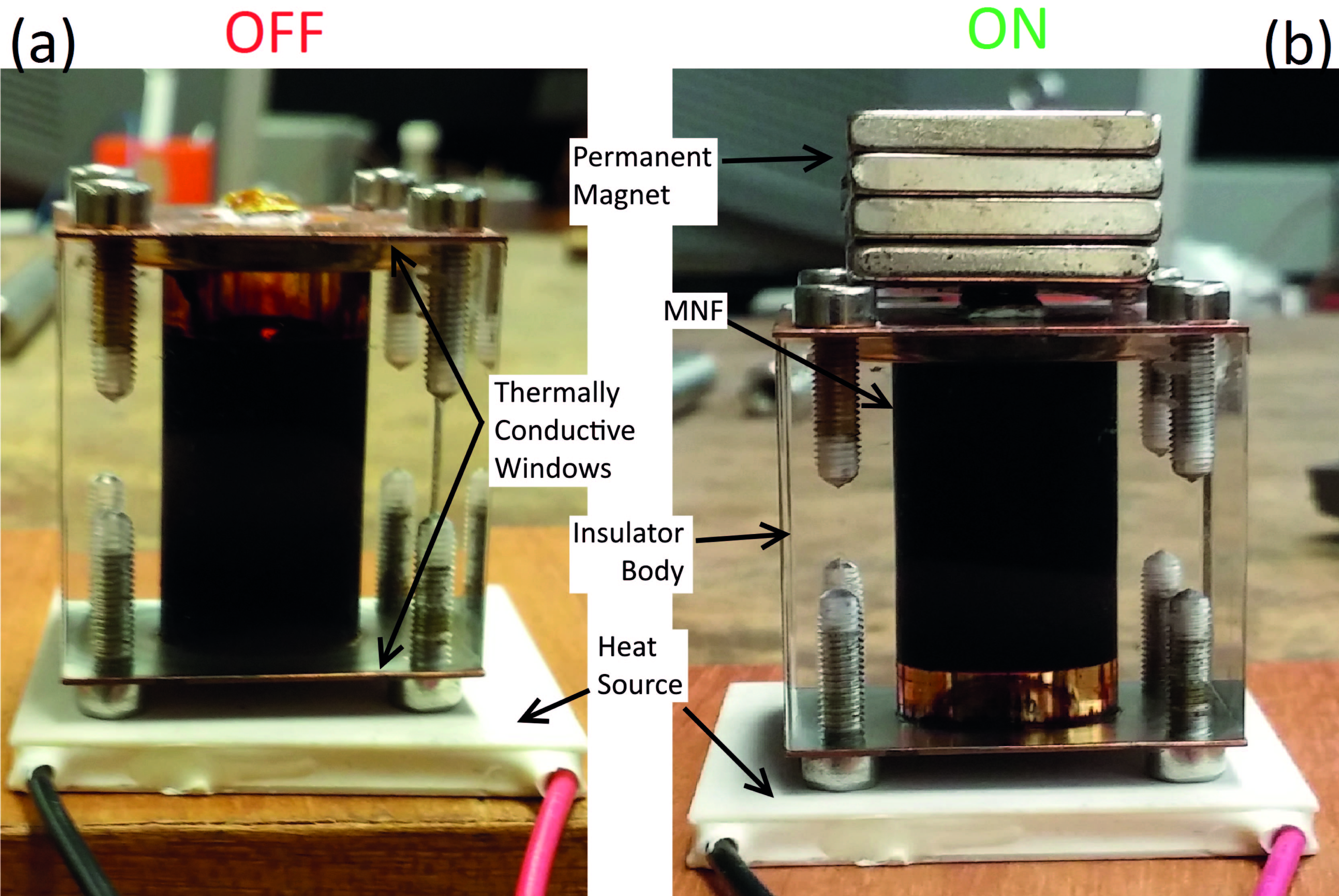}%
    \caption{Magnetically-Activated Thermal Switch prototype. (a) In the OFF state the nanofluid captures heat from the heat source and (b) transports it to the cold side as the magnetic field is applied.}
    \label{TS2}
\end{figure}

The MATS consists on a poly(methyl methacrylate) (PMMA) cubic container, with $3$ cm of side length and a cylindrical cavity in its center with $1.5$ cm of diameter (Fig. \ref{TS2}) partially filled with ferrofluid (approximately $4.42$ cm\textsuperscript{3}) and sealed with two copper plates. A small air spacer of approximately $0.5$ cm is left between the fluid and the top contact. The top copper plate is originally kept at room temperature while the bottom is heated with a Peltier element (Multicomp), creating a controllable temperature gradient.

\begin{figure}[!t]
    \centering
    \includegraphics[width=2.5in]{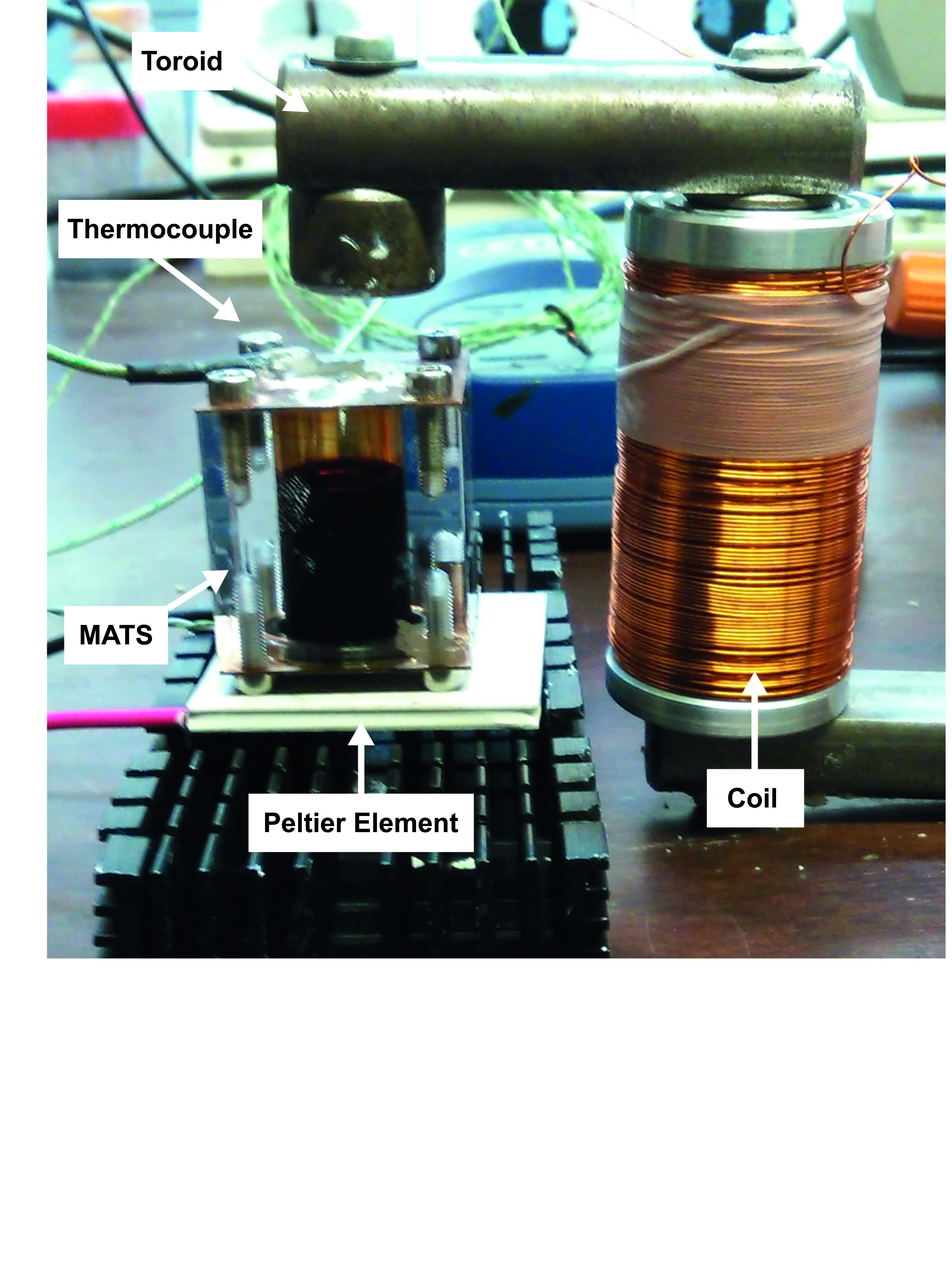}
    \caption{Experimental set-up used to test the thermal switch concept without mechanical parts.}
    \label{fig:2}
\end{figure}

An experimental set-up without movable parts was designed to test the thermal switch concept for various frequencies, ranging from $0.5$ to $18$ Hz. 
An electromagnet with a half horseshoe shape was assembled by placing a soft ferrite core partially inside a coil consisting of insulated copper wires (Fig. \ref{fig:2}).
When an electrical current flows through the copper coil, the ferrite material is magnetized and an external magnetic field (\textit{H}) is created.
When the current is turned on, the fluid rises to the upper contact due to the gradient of the magnitude of the applied magnetic field, according to \cite{coey_2010}:
\begin{equation}
\vec{F}= \nabla( \mu_{0} \vec{H} . \vec{M}),
\label{eq:LE}
\end{equation}
where $\vec{F}$ is the resulting force, $\mu_{0}$ is the magnetic permeability of vacuum, $\vec{H}$ is the external field and $\vec{M}$ the magnetization of the particles.
Then, when the current is turned off, the ferrofluid returns to its original position at the bottom of the container.

The current is applied to the coil in pulses using a waveform generator and a power supply. 
To control the applied current, a MOSFET was used as switch.
The temperatures on both cold and hot sides are recorded using two thermocouples. 
Each of the thermocouples was attached to a copper plate using a thin sheet of Indium and GE Varnish (GE-7031), to ensure an optimal thermal contact. 
The data is collected using a data logger (Pico Technology) and a custom made LabView program.

The FER-01 nanofluid used in the experiments with different temperature gradients is a commercially acquired nanofluid (Supermagnet) with a concentration of $4.4$ vol.\% of $10$ nm Fe$_{3}$O$_{4}$ nanoparticles dispersed in a poly-$\alpha$-olefin oil. 

\section{Experimental Results}

\subsection{MATS Operation}

\begin{figure}[!t]
    \centering
    \includegraphics[width=6in]{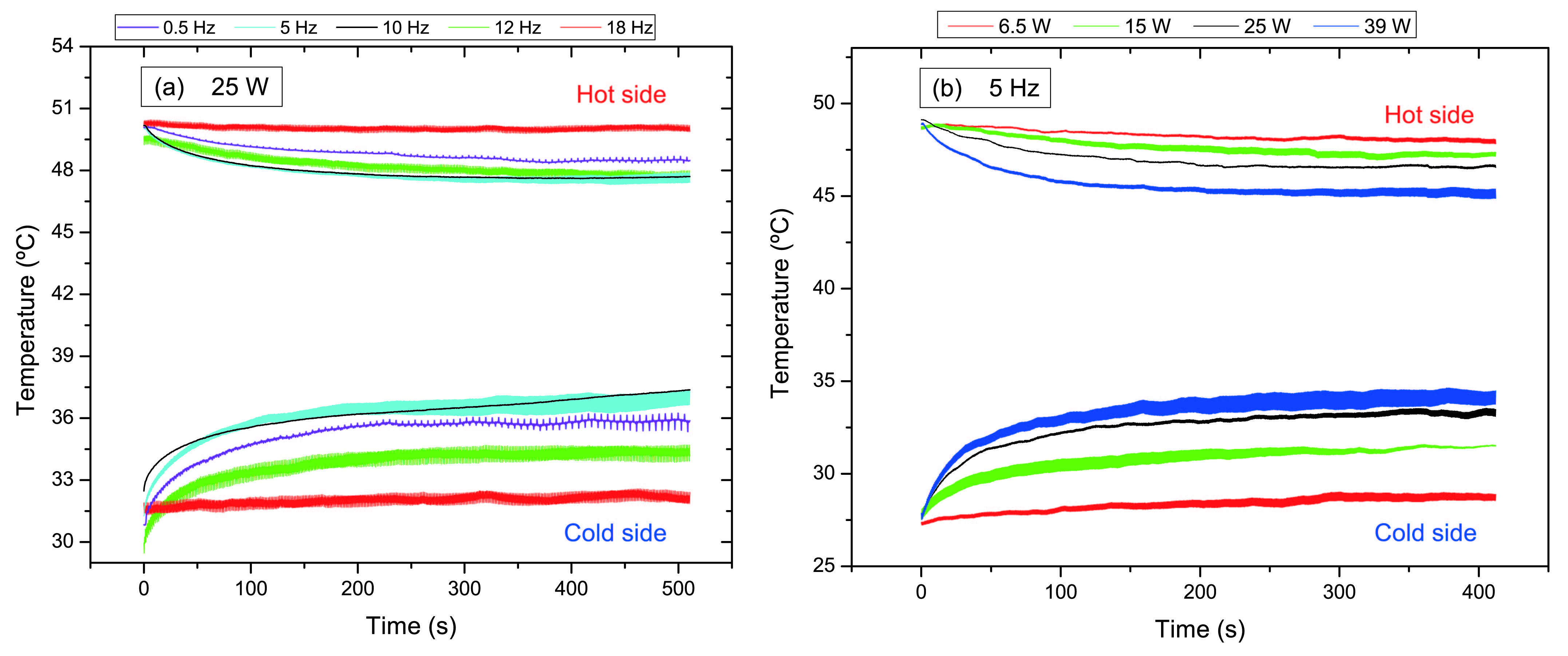}
    \caption{Experimental results for the FER-01 comercial nanofluid at $\Delta T = 20$ $^{\circ}$C for various (a) frequencies of applied magnetic field at 25 W and (b) powers applied in the coil at 5Hz.}
    \label{fig:4}
\end{figure}

The MATS was designed with the goal of removing heat from the hot side to the cold side by applying a magnetic field in a coil with a given frequency and without movable parts. 
At the start of the experiment, the temperature gradient across the MATS is stable at approximately 20 $^{\circ}$C [Fig. \ref{fig:4}(a)]. 
Then, at the $0$ s mark, the magnetic field is applied. 
As current flows through the coil, magnetizing the ferrite core, the magnetic nanofluid is pulled from the hot to the cold side, increasing the temperature of the later due to the transfer of heat from the hotter fluid to the colder surface. 
At the same time, the hot side also suffers from a temperature increase due to having less fluid in contact with the surface. 
When the current (and thus the magnetic field) is removed and the fluid returns to its original position, the temperature of the hot side suddenly decreases as the now colder fluid absorbs the heat generated by the Peltier. 
This process repeats itself with each cycle of the magnetic field leading to an overall decrease in temperature in the hot side and increase in the cold side.


Figure \ref{fig:4}(a) shows these results for the FER-01 commercial nanofluid for different frequencies. 
Generally speaking, the larger changes in temperature occur when the magnetic field is turned on. 
At low frequencies, the change in temperature for each individual cycle is small. For higher frequencies (between 5 and 12 Hz), the results show a higher decrease of the temperature gradient with the cold side experiencing a larger change (up to $6$ $^{\circ}$C) than the hot side (up to $2$ $^{\circ}$C), i.e, the higher frequency of the magnetic field allows a higher number of heat exchange cycles between the two surfaces.
However, further increasing the frequency to 18 Hz leads to a negligible decrease of the temperature gradient, as the time the nanofluid spends in the hot and cold sides is insufficient for a proper heat transfer to occur.   

Figure \ref{fig:4}(b) shows how different applied electrical powers affect the heat exchange between the hot and cold surface, with a fixed frequency of 5 Hz.
One can see that, with the increase of the power applied to the coil, the heat exchanges between the two surfaces become more efficient and the temperature gradient tends to decrease. 
For an applied power of 6.5 W one obtains a $\Delta$T of 19 $^{\circ}$C, while for 39 W, $\Delta$T decreases to only 10 $^{\circ}$C [Fig. \ref{fig:4}(b)].
However, such efficiency increase occurs at the cost of increased power consumption and thus a trade-off may be necessary for some applications.

\begin{figure}[!t]
    \centering
    \includegraphics[height=2.5in]{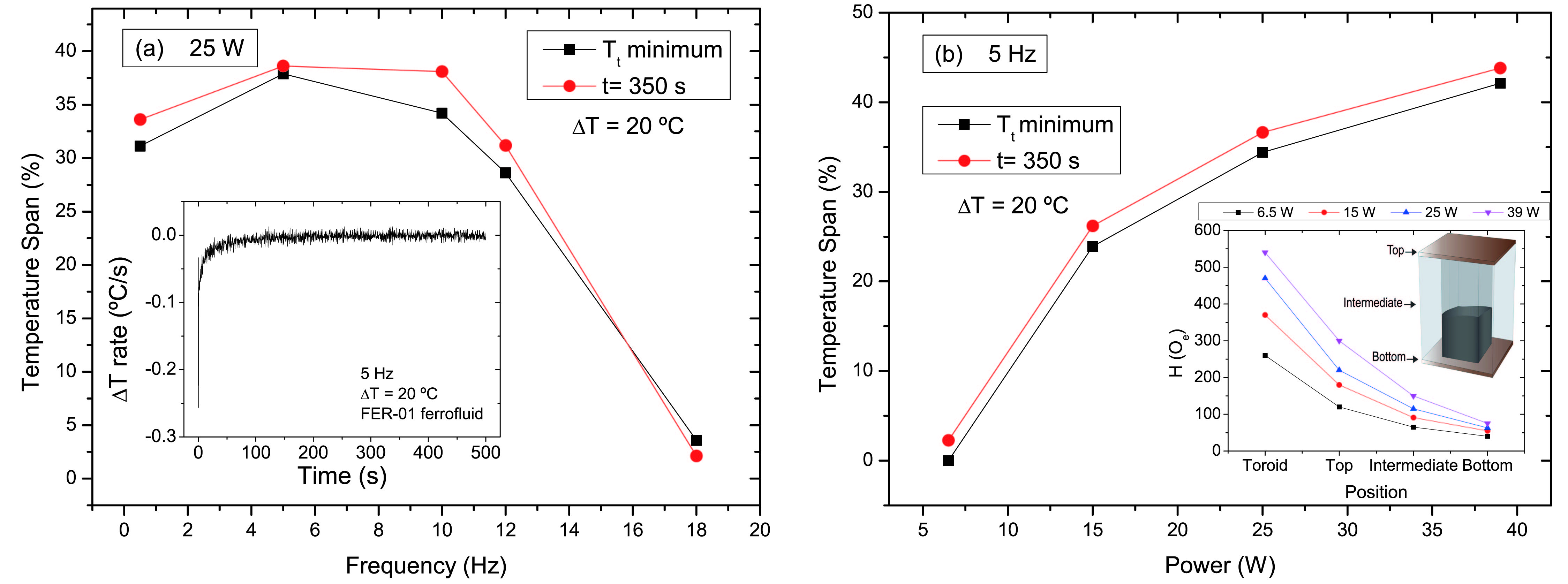}
    \caption{Temperature span for the FER-01 nanofluid at $t=350$ s and for $T_{t}$ minimum for the temperature gradients of 20 $^{\circ}$C: (a) for different frequencies  (the inset is the $\Delta T$ rate for the FER-01 nanofluid over time at $5$ Hz for $\Delta T=20 ^{\circ}$C.) and (b) various applied powers at 5 Hz (the inset is the magnetic field intensity of different positions of the cage for different applied powers).}
    \label{fig:5}
\end{figure}

To better evaluate the performance of the TS one must calculate how much the temperature gradient changes with the ON/OFF cycling of the magnetic field. 
For this we define the temperature span between the initial temperature gradient and the gradient at a given point in time ($T_{\%}$) as:

\begin{equation}
T_{\%}=(1-\frac{T_{t}}{T_{0}})\times 100\%,
\label{eq:DT}
\end{equation}
where $T_{t}$ is the temperature difference at $t$ seconds and $T_{0}$ is the temperature difference at $t=0$ s. 
A larger $T_{\%}$ indicates a larger decrease of the temperature gradient when compared to the initial gradient. 
This equation was used to evaluate the temperature span at i) $t=350$ s where, depending on the frequency, the temperature just entered or is already in the stable regime, and ii) when $T_{t}$ is minimum [Fig. \ref{fig:5}(a)]. 
In Figure \ref{fig:5}(a), we can see that the temperature span increases with increasing frequency until it reaches a maximum between 5 and 10 Hz, decreasing then for the higher frequencies. 
As the trends of the calculated temperature spans for each $T_{t}$ minimum and $t=350$ s are similar, in the following we will only consider the results of the former.

The largest changes in temperature occur just when the MATS starts operation. 
From the results, one can see that this change occurs at different rates ($\Delta T/ \Delta t$) for different experimental parameters. 
The inset of Fig. \ref{fig:5}(a) shows the variation of temperature over time ($\Delta T$ rate) for the FER-01 ferrofluid at $5$ Hz for an initial $\Delta T=20$ $^{\circ}$C.
When the magnetic field is applied (at $t=0$ s), there is a peak in the $\Delta T$ variation, with the temperature gradient decreasing at a rate of 0.26 $^{\circ}$C/s, followed by a gradual $\Delta T/ \Delta t$ decrease to values close to zero.

On the other hand, we can see that the temperature span increases with increasing power applied to the coil [Fig. \ref{fig:5}(b)].
By increasing the applied power, the magnetic force increases and consequently the nanofluid is more rapidly accelerated. 
This way, the higher the generated force, the larger the amount of nanoparticles that participate in the heat transport, thus making the MATS more efficient.
The inset of Fig. \ref{fig:5}(b) shows the intensity of the magnetic field measured at different MATS positions.
As expected, when applying the current pulse there is a magnetic field gradient along the MATS cage.
The H-magnitude is maximum near the half horseshoe shape (540 O$_{e}$ for an applied power of 39 W), while at the bottom side of the device H is minimum (75 O$_{e}$ for 39 W).


\subsection{Influence of $\Delta T$ on the Thermal Switch performance}

\begin{figure}[!t]
    \centering
    \includegraphics[width=6.5in]{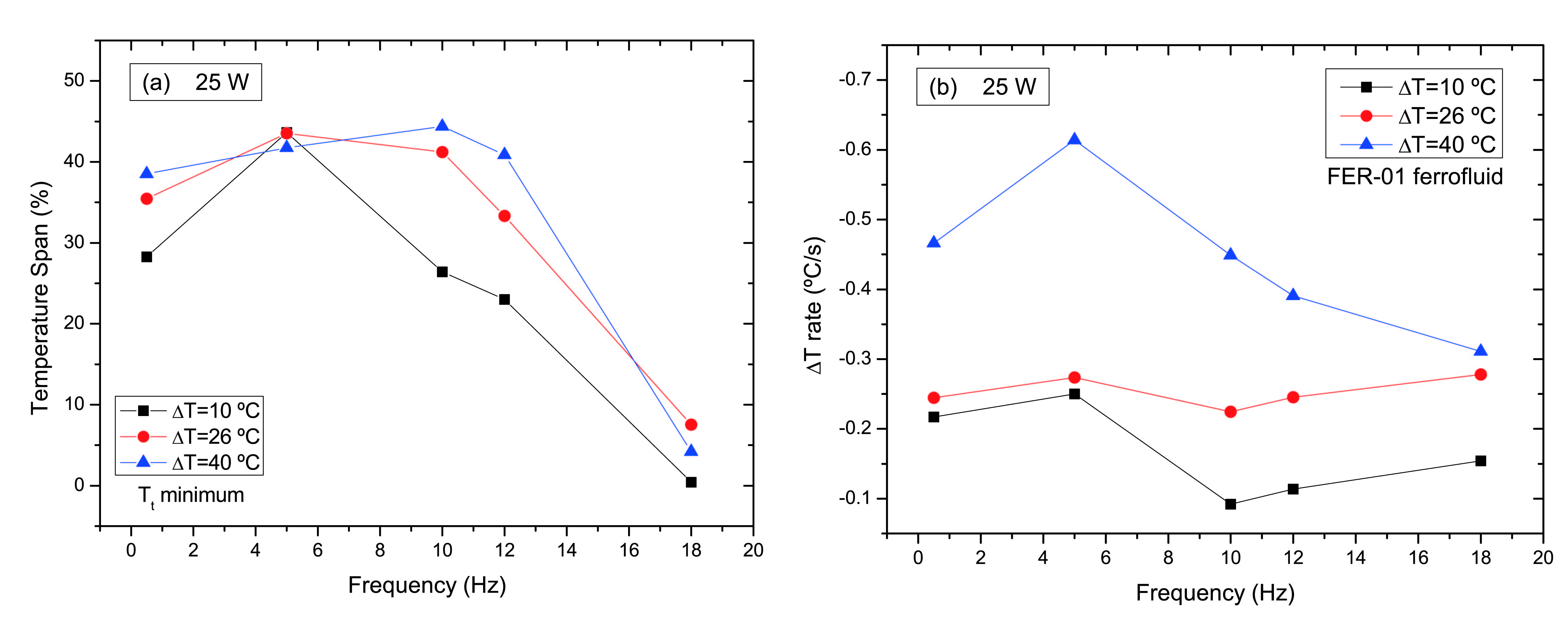}
    \caption{(a) Temperature span for the FER-01 nanofluid for $T_{t}$ minimum for the temperature gradients of $10$, $26$ and $40$ $^{\circ}$C. (b) Maximum $\Delta T$ rate for the three temperature gradients as a function of frequency.}
    \label{fig:7}
\end{figure}

Previous experiments with $\Delta T = 35$ $^{\circ}$C showed that frequency plays a major role in the performance of the TS \cite{Puga2017}. 
However, several applications work at lower temperature gradients and we thus varied the temperature gradient to evaluate its effects on the TS performance.
Figure \ref{fig:7} shows the temperature span as a function of frequency for $T_{t}$ minimum for three different temperature spans, $\Delta T =10$, $26$ and $40$ $^{\circ}$C. 
All show that the temperature span increases rapidly with frequency until it reaches a maximum between $5$ and $10$ Hz. 
For a frequency of $18$ Hz the temperature span decreases significantly for the three temperature gradients, as the fluid does not have time at each side of the TS to capture and release enough heat to maintain the performance. 
The MATS performance is overall best for the largest temperature gradient, $\Delta T = 40$ $^{\circ}$C. 
The maximum values obtained were of $43.6\%$ for $\Delta T=10$ $^{\circ}$C at 5 Hz, $43.5\%$ for $\Delta T=26$ $^{\circ}$C at 5 Hz and $44.4\%$ for $\Delta T = 40$ $^{\circ}$C at 10 Hz. 
However, the proximity of the temperature span values between temperature gradients in the low frequency region shows that the difference of operational temperatures does not greatly influence the performance of the MATS.
Comparing the $\Delta T/ \Delta t$ results for the three temperature gradient at different frequencies, one can see that, with increasing temperature gradients, the maximum $\Delta T/\Delta t$ magnitude also increases [Fig. \ref{fig:7}(b)]. 
Also, while $\Delta T=10$ $^{\circ}$C and $\Delta T=26$ $^{\circ}$C showed a small rate variation with frequency, $\Delta T = 40$ $^{\circ}$C presents larger changes. 
For $\Delta T = 40$ $^{\circ}$C, the $\Delta T$ rate increases significantly and presents a maximum for $5$ Hz.

\section{Conclusion}

We presented a novel MATS without moving parts that opens a new range of applications for this type of systems.
We showed that the magnetic field frequency, operation power and temperature gradient affect the performance of the MATS prototype.
For different temperature gradients, the MATS presents a slightly higher performance for $\Delta T = 40$ $^{\circ}$C (maximum $T_{\%}=44.4\%$). 
We also show that the temperature gradient rate increases with the initial $\Delta T$, from a maximum of $0.25$ $^{\circ}$C/s at $\Delta T=10$ $^{\circ}$C to $0.6$ $^{\circ}$C/s at $\Delta T= 40$ $^{\circ}$C.
It was found that, by increasing the magnetic force gradient, heat transport by the nanofluid becomes more efficient.
Therefore, the temperature span was maximum when the pulsed current was applied with a frequency of 10 Hz and 25 W. 
With these results we conclude that the use of magnetic nanofluids in a thermal switch is a viable option, showing a high performance even in early stages of development, leading to a broad range of applications as in magneto-caloric refrigeration \cite{Silva2012,Silva2014}.

\section*{Acknowledgments}

The authors acknowledge funding from FEDER and ON2 through project Norte-070124-FEDER-000070 and from FCT through the Associated Laboratory - IN. J. V. acknowledges financial support through FSE/POPH and project PTDC/CTM-NAN/3146/2014. This work was funded through project EXPL/EMS-ENE/2315/2013 from Fundacao para a Ciencia e Tecnologia (FCT), Portugal. C. Rodrigues thanks inanoEnergy for the research grant. B. D. Bordalo acknowledges FCT for grant PD/BD/114455/2016.


\bibliographystyle{model3-num-names}
\bibliography{bibliography}

\end{document}